\documentclass{PoS}
\usepackage{tabularx}
\usepackage{array}
\usepackage{graphics}
\usepackage{graphicx}
\usepackage{psfrag}
\usepackage{epsfig}
\usepackage{amsmath}
\usepackage{amssymb}
\usepackage{ulem}
\usepackage{setspace}
\usepackage{rotating}
\usepackage{colortbl}
\usepackage{tabularx}
\usepackage{longtable}
\usepackage{multirow}

\newcommand{\msbar}{{\overline{\rm MS}}}

\newcommand{\bea}{\begin{eqnarray}}
\newcommand{\eea}{\end{eqnarray}}
\newcommand{\beq}{\begin{equation}}
\newcommand{\eeq}{\end{equation}}
\newcommand{\ec}{\end{center}}
\newcommand{\bc}{\begin{center}}
\newcommand{\gev}{{\rm GeV}}
\newcommand{\mev}{{\rm MeV}}

\newcommand{\pdir}{p\kern -5.2pt\raise 0.2ex\hbox {/}}

\newcommand{\vdir}{v\kern -5.75pt\raise 0.15ex\hbox {/}}
\newcommand{\kdir}{k\kern -5.75pt\raise 0.15ex\hbox {/}}
\newcommand{\epsdir}{\epsilon\kern -5.0pt\raise 0.15ex\hbox {/}}
\newcommand{\bvdir}{\bar{v}\kern -5.75pt\raise 0.15ex\hbox {/}}
\newcommand{\Ddir}{D\kern -7.75pt\raise 0.20ex\hbox {/}}
\newcommand{\Adir}{A\kern -7.75pt\raise 0.20ex\hbox {/}}
\newcommand{\ldir}{l\kern -5.0pt\raise 0.2ex\hbox{/}}
\newcommand{\varepsdir}{\varepsilon\kern -5.5pt\raise 0.15ex\hbox{/}}

\newcommand{\nn}{\nonumber}

\title{ Radiative decays of charmonia on the lattice }

\ShortTitle{Radiative decays of charmonia on the lattice }

\author{\speaker{Francesco Sanfilippo}\\
  E-mail: \email{francesco.sanfilippo@th.u-psud.fr}}
\author{Damir Becirevic\\
  \\
  E-mail: \email{damir.becirevic@th.u-psud.fr}\\
  \\
  Laboratoire de Physique Th\'eorique (B\^at.~210)
  \footnote{Laboratoire de Physique Th\'eorique est une unit\'e mixte de recherche du CNRS, UMR 8627.}\\
  Universit\'e Paris Sud, F-91405 Orsay-Cedex, France.\\
  }

\abstract{We present the results of our lattice QCD study of the hadronic matrix elements relevant to the physical
radiative $J/\psi\to \eta_c\gamma$ and $h_c\to \eta_c\gamma$ decays. From computations with $N_{\rm f}=2$ dynamical
quark in twisted mass QCD at four lattice spacings, we were able to take the continuum limit and obtained 
$\Gamma(J/\psi \to \eta_c\gamma) = 2.64(11)~{\rm keV}$ and $\Gamma(h_c \to \eta_c\gamma) = 0.72(5)~\mev$.
We also computed the the hyperfine splitting and found that it does not depend from the sea quark mass and we obtain
$\Delta = m_{J/\psi}-m_{\eta_c} = 112\pm 4$~MeV.}

\FullConference{Xth Quark Confinement and the Hadron Spectrum,\\
		October 8-12, 2012\\
		TUM Campus Garching, Munich, Germany}

\begin{document}

\section{\label{sec-0}Introduction}
The radiative decay of $J/\psi\to\eta_c \gamma$ has been subject of extensive theoretical and experimental studies since many years. 
The current experimental results quoted by PDG~\cite{PDG} is,
\bea
\Gamma(J/\psi \to \eta_c\gamma) = 1.58(37)~{\rm keV}.
\eea

This value is obtained after averaging two experimental results, namely $\Gamma(J/\psi \to \eta_c\gamma) =1.18(33)~{\rm keV}$ by 
Crystal Ball~\cite{CRYSTAL}, and the more recent value obtained by CLEOc 1.91(28)(3)~keV~\cite{CLEOc}. 
The currently running KEDR experiment~\cite{KEDR} instead suggests a larger value, 2.2(6)~keV. It is fair to say that 
the current experimental situation is unclear and dedicated charm experiment at BESIII is expected to clarify the situation.

Prior to 2012 the theoretical situation concerning prediction of $\Gamma(J/\psi \to \eta_c\gamma)$ was not better. Dispersive 
analysis of $\Gamma\left(\eta_{c}\rightarrow2\gamma\right)$ obtained an upper bound for the width 
$\Gamma\left(J/\psi\rightarrow\eta_{c}\gamma\right)<$ 3.2 keV~\cite{shifman}. Two different QCD sum rule 
calculations resulted in $\left(1.7\pm0.4\right)$ keV~\cite{alex} and $\left(2.6\pm0.5\right)$, keV~\cite{Beilin}.
An effective theory of non relativistic QCD found $\left(1.5\pm1.0\right)\,{\rm keV}$~\cite{nora}.
Lastly, two different potential quark model calculatios exist, predicting an even larger value for the decay width, 
${\rm 3.3~keV}$~\cite{Voloshin}, and ${\rm 2.85~keV}$~\cite{Eichten}.
To all of these predictions the error must be regarded not as an estimate of uncertainty intrinsic to the method,
but only as susceptibility of the method to the variation of external parameters entering the predicition.
The global picture of the theoretical predictions for the $\Gamma(J/\psi \to \eta_c\gamma)$ is 
puzzling and inconclusive, and necessities a fully non perturbative analysis from the first principles of QCD.

The first extensive study of the radiative decays of charmonia on the lattice has been reported in ref.~\cite{lattice-radiative-1}
where the authors computed relevant matrix elements for a number of decay channels in the quenched approximation of QCD and with one
lattice spacing only. That computation has been extended to the case of $N_{\rm f}= 2$ dynamical light quark flavors at single lattice
spacing in ref.~\cite{lattice-radiative-2}. In this paper we will focus on $J/\psi \to \eta_c\gamma$ and $h_c\to \eta_c\gamma$, for which
we compute the desired form factors for four lattice spacings that we could extrapolate to the continuum limit. 
Our result for $\Gamma(J/\psi \to \eta_c\gamma)$ allow for a clear comparison between theory and experiment, as soon as the
results from KEDR and BESIII become available. Our $\Gamma(h_c\to\eta_c\gamma)$ will provide us with a prediction for $h_c$ lifetime, 
and both could be compared with experimental measure when it becomes available.

This presentation is based on our recent paper~\cite{papero} where the interested reader can find details of all the
numerically computed data entering our calculation.

\section{\label{sec-2}Hadronic Matrix Elements }

The transition matrix element responsible for the $J/\psi\to \eta_c\gamma^\ast$ decay reads, 
\bea\label{def-vectorFF}
\langle \eta_c(k) \vert  J^{\rm em}_\mu \vert J/\psi(p,\epsilon_\lambda) \rangle = 
e {\cal Q}_c\ \varepsilon_{\mu\nu\alpha\beta}\  \epsilon_\lambda^{\ast \nu} p^\alpha k^\beta \ 2\ V(q^2)/(m_{J/\psi}+m_{\eta_c})\,,
\eea
where $J^{\rm em}_\mu ={\cal Q}_c \bar c \gamma_\mu c$ is the relevant piece of the electromagnetic current, with ${\cal Q}_c=2/3$ 
in units of $e=\sqrt{4\pi \alpha_{\rm em}}$. Information regarding the non-perturbative QCD dynamics is encoded in the form factor 
$V(q^2)$ and represents the most challenging part on the theory side.  For the physical process, i.e. with the photon on-shell
$q^2=0$, the decay rate is given by~\cite{lattice-radiative-1}
\bea\label{widthPSI}
\Gamma(J/\psi\to \eta_c\gamma)& = & {8\over 27}\ \alpha_{\rm em}\   (m_{J/\psi}+m_{\eta_c} )\  \left({\Delta \over m_{J/\psi}}\right)^3 \left| V(0)\right|^2\,,\quad\Delta=m_{J/\psi}-m_{\eta_c}\,.
\eea
Similarly $h_c\to \eta_c\gamma$ transition matrix element is parametrized by two form factors $F_{1,2}(q^2)$,
\bea\label{def-vectorFH} 
{\langle \eta_c(k) \vert  J^{\rm em}_\mu \vert h_c(p,\epsilon_\lambda) \rangle \over  i e {\cal Q}_c} = m_{h_c} F_1(q^2) \left(  \epsilon^{\lambda \ast}_\mu - { \epsilon_\lambda^\ast  q\over q^2} q_\mu \right)+ F_2(q^2) ( \epsilon_\lambda^\ast q) \left[
 { m_{h_c}^2 - m_{\eta_c}^2\over q^2} q_\mu  -  (p+k)_\mu \right]\,. 
\eea
The decay rate for the on-shell photon is~\cite{lattice-radiative-1}
\bea\label{widthH}
\Gamma(h_c\to \eta_c\gamma)=8\alpha_{\rm em} (m_h^2-m_{\eta_c}^2) \left| F_1(0)\right|^2/(27 m_{\eta_c}^2)\,.
\eea
We can compute the form factor $V(q^2)$ and $F_1(q^2)$ directly at $q^2=0$ by using twisted boundary 
conditions~\cite{twbc} on one propagator, which will be labelled in the following by a superscript ``$\theta$''.

\section{Two-point correlation functions}
As in~\cite{Becirevic:2012ti}, we use the maximally twisted mass QCD~\cite{fr} gauge field configurations produced by ETM 
collaboration~\cite{boucaud}.
We extract mass of charmonia from the two point correlation functions:
\bea\label{eq:0}
 C^\Gamma(t) = \langle \sum_{\vec x} {\rm Tr} \left[ O^\Gamma(\vec 0,0) O^\Gamma(\vec x,t)\right] \rangle, \quad\quad O^\Gamma(x)=\bar{c}(x)\Gamma c(x)\,,
\eea
in which the Dirac structures $\Gamma$ are chosen as $\gamma_5, \gamma_i$ or $\sigma_{0i}, i\in(1,2,3)$ to provide the coupling to 
the charmonium states with quantum numbers $J^{PC}=0^{-+}$, $1^{--}$, and $1^{+-}$, for $\eta_c$, $J/\psi$ and $h_c$, respectively. 
We implement the Gaussian smearing on the fermionic fields $c$ entering \ref{eq:0}, and compute the quark propagators using stochastic 
techniques \cite{boucaud}.
Charm quark mass, $\mu_c$, at each of our lattices 
has been fixed according to the result of ref.~\cite{Blossier:2010cr} where it was shown that the charm quark computed from the 
comparison of the lattice results with the physical $m_{\eta_c}$ fully agrees with the value obtained by using the physical 
$m_{D_s}$ or $m_{D}$. Therefore, we can say that $m_{\eta_c}$, obtained by computing the effective mass $m_{\eta_c}^{\rm eff}$ 
from correlation function $C^{\eta_c}(t)$,
\begin{figure}[t!!]
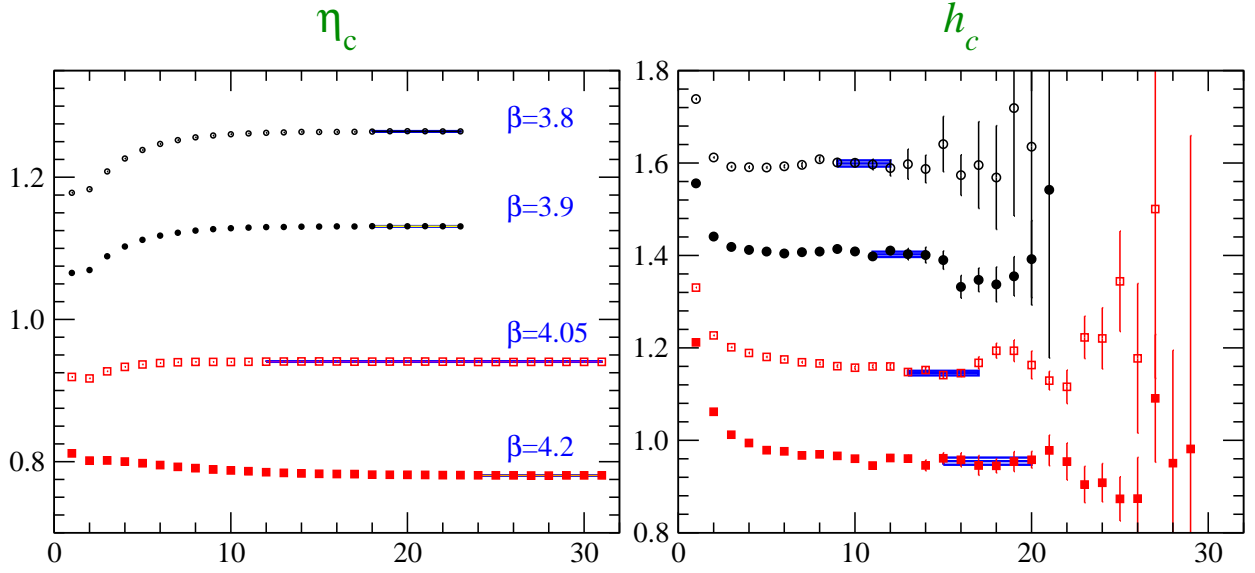

\begin{center}
\begin{tabular}{@{\hspace{-0.25cm}}c}
\epsfxsize8.2cm\epsffile{fig1a.eps} \epsfxsize8.2cm\epsffile{fig1c.eps}   \\
\end{tabular}
\vspace*{-.1cm}
\caption{
  \label{fig:1}
        {\footnotesize
          Effective masses of the charmonium states, $m_{\eta_c,h_c}^{\rm eff}(t)$, extracted from the two-point correlation
          functions at four lattice spacings, for one value of the sea quark mass.}
}
\end{center}
\end{figure}
and then by fitting $m_{\eta_c}^{\rm eff}(t)$ at large time separations to a constant, is merely  a verification that, after a 
smooth continuum extrapolation, we indeed reproduce $m_{\eta_c}^{\rm exp.}=2.980(1)$~GeV. To extract the values of 
$m_{J/\psi}$ and $m_{h_c}$ we proceed along the same line, computing  $m_{J/\psi,h_c}^{\rm eff}(t)$ using the
appropriate correlation function, and then fitting them at sufficientely large time separation.

In fig.~\ref{fig:1} we show an example of two effective mass plots, as obtained by using all four lattice spacings explored in
this work and for one value of the sea quark mass, which we choose to be the least light ones, for the case of $\eta_c$ and of
$h_c$ mesons. We see that the effective masses for the pseudoscalar are excellent while the signal for $h_c$ is  good but more noisy.
The quality in the case of $J/\psi$ (not shown) lies between the two illustrated. The effective masses are then combined to
\bea\label{eq:RRR}
R_{J/\psi}(t)= {m_{J/\psi}^{\rm eff}(t)\over m_{\eta_c}^{\rm eff}(t)}\,,\quad 
R_{h_c}(t)= {m_{h_c}^{\rm eff}(t)\over m_{\eta_c}^{\rm eff}(t)}\,.
\eea
We then fit $R_{J/\psi,h_c}(t)$ plateaus to a constant $R_{J/\psi,h_c}$, extrapolating to the continuum limit, we get
\bea
\label{eq:contR}
R_{J/\psi,h_c} = R_{J/\psi,h_c}^{\rm cont.} \left[ 1 +  b_{J/\psi,h_c} m_{q} + c_{J/\psi,h_c} a^2/a_{3.9}^2\right]\,,\quad\quad\quad\quad\quad\\
\nn\\
\label{eq:resR}
R_{J/\psi }^{\rm cont.}  = 1.0377(6)\quad[\ 1.0391(4)\ ]^{\rm exp.}\,,\quad\quad
 R_{ h_c}^{\rm cont.} = 1.187(11)\quad [\ 1.1829(5)\ ]^{\rm exp.}\,.
\eea
In eq.~(\ref{eq:contR}) the 
parameter $b_{J/\psi,h_c}\approx 0$ measures the dependence on the sea quark mass, $m_q\equiv m_q^\msbar(2\ \gev)$, while 
the parameter $c_{J/\psi,h_c}\approx 3$~\% measures the leading discretization effects. Division by $a_{\beta=3.9}=0.086$~fm 
is made for convenience. The linear fit~(\ref{eq:contR}) describes our data very well except for the results obtained at 
$\beta=3.8$, that can be either excluded from the extrapolation (above results), or included adding a term proportional to $a^4$,
leading to a fully consistent result with the one quoted above. Having neglected the disconnected
contributions to the correlation functions, the fact that our lattice results agree with the experimental values (\ref{eq:resR})
can be viewed as a verification that they are indeed very small.

From eq.~(\ref{eq:resR}), one can infer the hyperfine splitting:
\bea
\Delta = m_{J/\psi}-m_{\eta_c} = m_{\eta_c}( R_{J/\psi}-1) \,, 
\eea
which after linar fit to a pametrization similar to that of eq.(~\ref{eq:contR}) gives
\bea\label{latt-D}
\Delta^{\rm cont.} = (112\pm 4)~\mev&&\quad[\ 116.6\pm 1.2\ ]^{\rm PDG.}\,,
\eea
in good agreement with the experimental result written in brackets~\cite{PDG}, and in excellent agreement with the result of 
BESIII~\cite{Ablikim:2010rc}, $\Delta = (112.4\pm 1.16)~\mev$.
Note also that from the fit of our data we find $\Delta =1.0(3)\ \gev^{-1}$, in qualitative agreement with 
ref.~\cite{0912} where a tiny decrease of $\Delta$ is found while lowering the sea quark mass.
Note, however, that this observation ($b_\Delta \gtrapprox 0$) disagrees with earlier findings of ref.~\cite{Manke:2000dg}.

\section{\label{sec-3}Radiative Transition Form Factors}

\begin{figure}
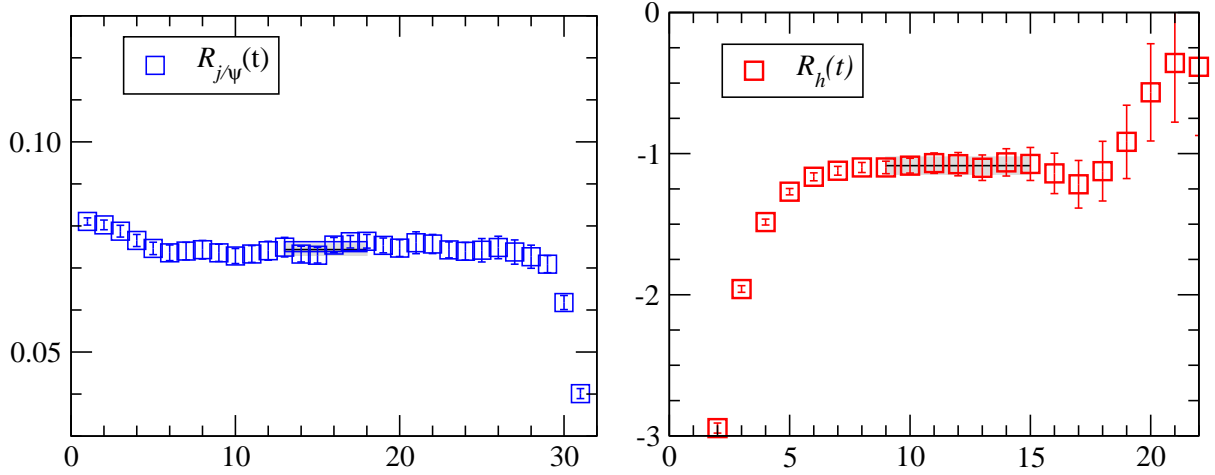

\vspace*{-0.8cm}
\begin{center}
\begin{tabular}{@{\hspace{-0.25cm}}cc}
\epsfxsize7.9cm\epsffile{fig6b.eps}
&\epsfxsize7.7cm\epsffile{fig6d.eps}
\end{tabular}
\vspace*{-.1cm}
\caption{\label{fig:5}{\footnotesize
    Plateaus exhibited by $R_{J/\psi}(t)$ and $R_{\rm h}(t)$ for the heaviest sea quark at $\beta=4.05$.
} }
\end{center}
\vspace*{-0.7cm}
\end{figure}

To extract the desired hadronic matrix element~(\ref{def-vectorFF}) we computed the three point correlators
\bea\label{eq:c31}
C_{ij}(\vec q;t) = \sum_{\vec x,\vec y} \langle V_i^\dag (0) J^{\rm em}_j(x) P(y) \rangle \ e^{i\vec q\cdot (\vec x -\vec y )} 
= \langle \sum_{\vec x,\vec y}{\rm Tr}\left[  S_c (y;0) \gamma_i S_c (0,x)\gamma_j   S_c^{ \vec \theta}(x,y) \gamma_5 \right]\rangle \,,
\eea
where $P=\bar c\gamma_5 c$, $V_i =\bar c\gamma_i c$ are the interpolating operators fixed at $t=0$ and $t=t_y=T/2$ ($T$ being the 
time extension of our lattices). Using the fact that our three-momentum $\vec q=\vec\theta/L$ is isotropic, we 
averaging 6 equivalent contributions, $C_{12}$, $C_{23}$, $C_{31}$, $-C_{21}$, $-C_{32}$, $-C_{13}$ to $C_V$,
\bea\label{eq:vvv}
C_V(\vec q;t) \to {{\cal Z}_P^S {\cal Z}_V^S \over 2 E_{\eta_c} }\, {\theta_0\over L}\, {e^{-[E_{\eta_c}t_{\rm fix}+(m_{J/\psi}-E_{\eta_c})t]}\over m_{J/\psi} + m_{\eta_c}}\, V(0)
\,,
\eea
where the last expression is valid for sufficiently separated operators in the correlation function~(\ref{eq:c31}).
By renormalizing the local electromagnetic current and combining appropriately two and three 
points correlators, we can build a ratio $R_{J/\psi}$ where we eliminate the source terms, obtaining 
$V(0)$ from a fit to a constant as shown in fig.~\ref{fig:5}.

Extrapolation the physical limit ($m_{\rm sea}\equiv m_q\to 0$, $a\to 0$) is performed using a form similar to eq.(\ref{eq:contR}).
We do not observe any dependance of $V(0)$ on the light sea quark mass. Instead the discretization effects are rather large, 
with $c_V=-23\%$. Our final result is:
\bea\label{FINALV}
V(0)=1.94(4)\,.
\eea
\begin{figure}[h]
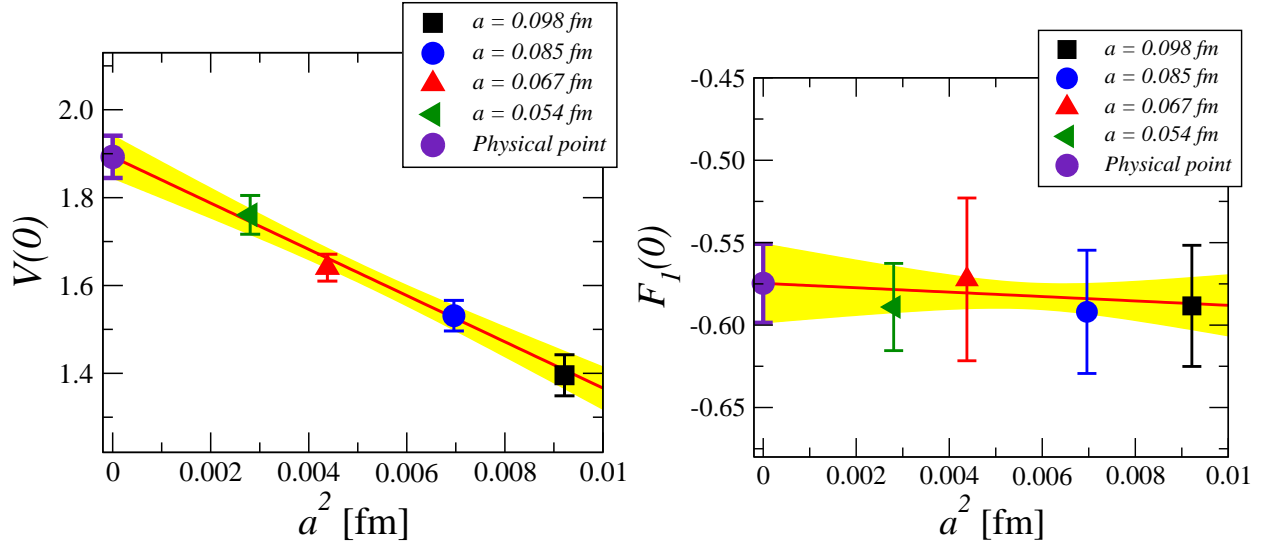

\begin{center}
\begin{tabular}{@{\hspace{-0.25cm}}c}
\epsfxsize8.2cm\epsffile{fig7a.eps}~\epsfxsize8.2cm\epsffile{fig7b.eps}   \\
\end{tabular}
\vspace*{-.1cm}
\caption{\label{fig:6}{\footnotesize 
    Linear continuum extrapolation of the form factors $V(0)$ and $F_1(0)$.
} }
\end{center}
\end{figure}
We now turn to the discussion of the form factor $F_1(0)$, relevant to the $h_c\to \eta_c\gamma$ decay, as defined in 
eq.~(\ref{def-vectorFH}). To that end we compute the three point correlators
\bea\label{Cijk}
C_{ijk} (\vec {q};t)= \sum_{\vec x,\vec y} \langle T_{ij}^\dag(0) J^{\rm em}_k(x) P(y) \rangle \ e^{i\vec {q}( \vec x- \vec y)} =  - \langle \sum_{\vec x,\vec y}{\rm Tr}\left[  S_c(y;0) \gamma_i \gamma_j S_c (0,x)\gamma_k  S_c^{ \vec{\tilde \theta} }(x,y) \gamma_5 \right]\rangle \,.
\eea
$F_1(0)$ is then obtained fitting to a constant the ratio obtained dividing the combination 
$C_{F_1} = \left[C_{123}+C_{231}+C_{312}\right]/3-\left[C_{131}+C_{212}+C_{323}+C_{232}+C_{313}+C_{121}\right]/6$ with the two point functions
(see fig.~\ref{fig:6}). We perform the continuum and chiral extrapolations, in a way analogous to eq.~(\ref{eq:contR}).
Again, the form factor $F_1(0)$ is insensitive to the variation of the light sea quark mass. Contrary to $V(0)$, the 
discretization effects turn out to be smaller: we find $c_{F_1}\approx 2$\%, and our final results is
\bea\label{FINALF1}
F_1(0) = -0.57(2)\,.
\eea
\section{\label{sec-5}Phenomenology }
\subsection{Decays of $J/\psi$}
Concerning the radiative decay $J/\psi \to \eta_c\gamma$, by inserting our value (\ref{FINALV}) in eq.~(\ref{widthPSI}) we get
\bea\label{eq:ourPSI}
\Gamma(J/\psi \to \eta_c\gamma) = 2.64(11)~{\rm keV}\qquad [1.58(37)~{\rm keV}]^{\rm exp.},
\eea
where we used the measured ${\rm Br}(J/\psi\to \eta_c\gamma)= (1.7\pm 0.4)\%$, the full
width $\Gamma_{J/\psi}= 92.9\pm 2.8$~keV~\cite{PDG}, the physical values of $m_{J/\psi}=3096.92(1)$~MeV 
and $\Delta=116.6\pm 1.2$~MeV. 

Our result for the decay rate is larger than the experimental one, and the agreement is only at $2\sigma$. 
The various effective approaches presented in the introduction agree with ours too, except that we have smaller and controlled 
uncertainty. We hope more effort on the experimental side will be devoted to clarify the disagreement among various experiments.

We note also that the quenched result of ref.~\cite{lattice-radiative-1}, 
$V(0)=1.85(4)$, is only slightly lower than ours, while the one obtained with $N_{\rm f}=2$ light flavors with a single lattice 
spacing in ref.~\cite{lattice-radiative-2}, is larger than ours at the same lattice spacing.

We do not make any estimate of the size of systematic uncertainty due to the omitted $s$ and $c$ quarks in the 
sea. Very recently a study similar to ours has been made in ref.~\cite{DAVIES} where several set of gauge field 
configurations, obtained with HISQ action including $N_{\rm f}=2+1$ dynamical quark flavors. Their results are in
perfect agreement with those reported here, which means that (i) the inclusion of the strange quark in the sea has no 
impact on $J/\psi\to\eta_c\gamma$ and (ii) that the continuum results obtained by two totally different 
lattice regularization lead to perfectly consistent values.

\begin{figure}
\begin{center}
\epsfxsize13cm\epsffile{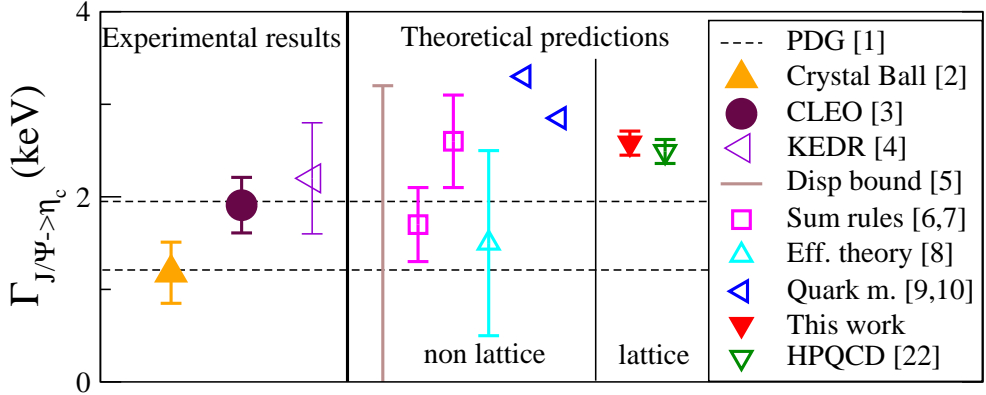}  \\
\caption{
  \label{fig:comparison_Davies}
        {\footnotesize Comparison of different results for $\Gamma(J/\psi \to \eta_c\gamma)$.
}
}
\end{center}
\vspace{-1cm}
\end{figure}

In fig.~\ref{fig:comparison_Davies} we present full comparison of the experimental and theoretical findings on 
$J/\psi \to \eta_c\gamma$.

\subsection{$h_c  \to \eta_c \gamma$}

$h_c$ escaped the experimental detection for a long time and only recently CLEO succeeded to isolate this state~\cite{Rosner:2005ry}
and observed that its prominent mode is precisely $h_c\to \eta_c\gamma$, the branching fraction of which was later accurately 
measured at the BESIII experiment, with a result: ${\rm Br}(h_c\to \eta_c\gamma)= (53\pm 7)\%$~\cite{Ablikim:2010rc}. We obviously 
cannot compute the branching ratio on the lattice, but with our form factor result~(\ref{FINALF1}) we can compute the decay width 
using eq.~(\ref{widthH}). We get
\bea
\Gamma(h_c \to \eta_c\gamma) = 0.72(5)~\mev\,.
\eea
This can be combined with the measured ${\rm Br}(h_c\to \eta_c\gamma)$ to estimate the width of $h_c$. We obtain:
\bea
\Gamma_{h_c}= {\Gamma(h_c \to \eta_c\gamma) \over {\rm Br}(h_c\to \eta_c\gamma)} = 1.37\pm 0.11\pm 0.18\ \mev\, = 1.37 \pm 0.22\ \mev,
\eea
where the first error comes from our determination of the form factor $F_1(0)$, and the second one reflects the experimental 
uncertainty in the branching ratio. This constitutes a prediction that will be interesting to check against experiment
once the latter becomes available.

To compare with other lattice results we convert the value reported in 
ref.~\cite{lattice-radiative-1} to our dimensionless form factor and obtain  $F_1(0)=-0.53(3)$, which agrees very well with our 
result. Similar conversion of the result of ref.~\cite{lattice-radiative-2} would result in $F_1(0)=-0.33(1)$, much smaller value than 
ours, whether we compare it with the values we obtain at $\beta=4.05$ or the one in the continuum. 

\section{\label{sec-6}Summary and future perspectives}

We presented results of our analysis of the radiative decays of charmonia by means of QCD simulations on the lattice. 
Using several lattice spacings of twisted mass QCD with $N_{\rm f}=2$ dynamical flavors we were able to 
extrapolate the relevant form factors to the continuum limit. 

We emphasize that our results are obtained without inclusion of disconnected diagrams.

\section*{Acknowledgments}
Computations are performed using GENCI (CINES) Grant 2012-056806.

\end{document}